\documentclass[twocolumn,aps]{revtex4}

\usepackage{graphicx} 
\usepackage{color} 

\begin{document}

\title{High threshold universal quantum computation on the surface code}

\author{Austin G. Fowler$^{1}$, Ashley M. Stephens$^{1}$ and Peter Groszkowski$^{2}$}
\affiliation{$^{1}$Centre for Quantum Computer
Technology,\\University of Melbourne, Victoria, AUSTRALIA\\
$^{2}$Institute for Quantum Computing, University of Waterloo,
Waterloo, ON, CANADA}

\date{\today}

\begin{abstract}
We present a comprehensive and self-contained simplified review of
the quantum computing scheme of \cite{Raus07,Raus07d}, which
features a 2-D nearest neighbor coupled lattice of qubits, a
threshold error rate approaching 1\%, natural asymmetric and
adjustable strength error correction and low overhead, arbitrarily
long-range logical gates. These features make it by far the best and
most practical quantum computing scheme devised to date.  We
restrict the discussion to direct manipulation of the surface code
using the stabilizer formalism, both of which we also briefly
review, to make the scheme accessible to a broad audience.
\end{abstract}

\maketitle

\section{Introduction}

Classical computers manipulate bits that can be exclusively 0 or 1.
Quantum computers manipulate quantum bits (qubits) that can be
placed in arbitrary superpositions $\alpha|0\rangle +
\beta|1\rangle$ and entangled with one another $(|00\rangle +
|11\rangle)/\sqrt{2}$. This additional flexibility provides both
additional computing power and additional challenges when attempting
to correct the now quantum errors in the computer. An extremely
efficient scheme for quantum error correction and fault-tolerant
quantum computation is required to correct these errors without
making unphysical demands on the underlying hardware and without
introducing excessive time overhead and thus wasting a significant
amount of the potential performance increase.

This paper is a simplified review of the quantum computing scheme of
\cite{Raus07,Raus07d}.  The scheme requires a 2-D square lattice of
nearest neighbor coupled qubits with initialization, readout, memory
and quantum gates all operating with error rates less than
approximately 1\% --- the least challenging set of physical
requirements devised to date. Furthermore, despite the modest
physical requirements, logical qubits (qubits of data distributed
over many physical qubits and protected by error correction) can be
interacted over arbitrarily large distances with time overhead only
growing logarithmically in their separation.  This is remarkable as
most nearest neighbor quantum computing schemes are associated with
a time overhead that grows linearly with logical qubit separation.
Finally, in most, if not all, physical quantum computer
technologies, bit-flips $|0\rangle \leftrightarrow |1\rangle$ are
less likely than phase-flips $|1\rangle \leftrightarrow -|1\rangle$
opening the door for asymmetric error correction schemes that make
use of fewer physical qubits to preserve a given amount of quantum
data with a given confidence level.  Practically, it is also helpful
if additional error correction resources can be dynamically
allocated to critical data during the quantum computation.  The
scheme we review permits both asymmetric and dynamic error
correction in a natural manner.

A number of technologies are well-suited to implementing surface
code quantum computing.  Proposals exist for 2-D arrays of qubits
making use of superconductors \cite{Helm09,DiVi09} and semiconductor
nanophotonics \cite{VanM09}. An equivalent measurement based version
of the scheme calling for a 3-D cluster state
\cite{Raus07,Raus07d,Fowl09} could be implemented using photonic
modules \cite{Devi07,Devi08} or ion traps \cite{Stoc08}.

The discussion is organized as follows.  In Section~\ref{Quantum
states and stabilizers} we briefly review the stabilizer formalism
of quantum computing \cite{Gott97}. Section~\ref{The surface code}
briefly reviews the surface code \cite{Brav98}, which forms the
error correction substrate of everything that follows.  Logical
qubits are introduced into the surface code in Section~\ref{Logical
qubits}, along with their initialization, measurement, and basic
logical operations. Section~\ref{Logical CNOT} describes logical
CNOT in detail. Section~\ref{State injection and non-Clifford gates}
completes the universal set of logical gates with a discussion of
state injection, state distillation and appropriate quantum circuits
making use of the distilled states.  An efficient implementation of
logical Hadamard inspired by \cite{Bomb07} that avoids the extensive
machinery of Section~\ref{State injection and non-Clifford gates} is
described in Section~\ref{Logical Hadamard}. Section~\ref{Threshold
error rate} then describes simulations used to estimate the
threshold error rates of physical qubit initialization, measurement,
memory and two-qubit gates. Looking further into the future,
Section~\ref{Distributed computing} discusses distributed quantum
computing to make it clear that impractically large 2-D square
lattices of qubits are not required to tackle problems of
interesting size. Section~\ref{Conclusion and further reading}
summarizes the discussion and points to further reading.

\section{Quantum states and stabilizers}
\label{Quantum states and stabilizers}

A quantum state can be specified in a number of equivalent ways. One
of the most common is to choose a basis and express the state as a
state vector such as $(|00\rangle + |11\rangle)/\sqrt{2}$.  In this
review, it will be much more convenient to express this state as the
unique simultaneous $+1$ eigenvector of the commuting operators
$X\otimes X$ and $Z\otimes Z$.  Such operators are called
stabilizers \cite{Gott97}. This entire review is based on the
manipulation of stabilizers.

Any set of $n$ mutually commuting and independent operators over $n$
qubits has a unique simultaneous $+1$ eigenstate. We will restrict
our attention to stabilizers that are a tensor product of the
identity operator $I$ and the Pauli matrices $X$, $Y$, $Z$ (with
$Y=XZ$ real).  A set of such stabilizers cannot be used to specify
an arbitrary quantum state, though a sufficiently broad range of
states can be specified for most of our purposes.  See
Section~\ref{State injection} for a simple extension to the
stabilizer formalism permitting arbitrary states to be specified.

Consider a set of stabilizers $M$ specifying state $|\psi\rangle$.
Suppose we wish to apply an operator $U$ to state $|\psi\rangle$. If
we consider $U|\psi\rangle = UMU^{\dag}U|\psi\rangle$, we can see
that the new set of stabilizers will be $UMU^{\dag}$. To give an
explicit example,
\begin{eqnarray}
M & = & Z_1Z_2, Z_2Z_3, Z_3Z_4, X_1X_2X_3X_4\\
U & = & X_2\\
UMU^{\dag} & = & -Z_1Z_2, -Z_2Z_3, Z_3Z_4, X_1X_2X_3X_4.
\end{eqnarray}

In addition to unitary manipulation, we will frequently discuss
measurement of a given operator, for example $X$, $Z$ or some more
complicated operator involving a larger tensor product.  A very
simple example is a single qubit in an unknown state with stabilizer
$I$ and the subsequent measurement of the $Z$ operator. We will
write the stabilizer of a qubit after such a measurement as $\pm Z$.
Note that the probabilities of the two possible measurement
outcomes, the $+1$ and $-1$ eigenstates of $Z$, are typically not
recorded in the stabilizer formalism, just their possibility.  Note
also that given any operator there is always a nonzero probability
of obtaining at least one of the two eigenstates.

Care needs be taken when measuring if other nontrivial stabilizers
are present.  There are three cases to consider.  If the operator to
be measured can be expressed as a product of stabilizers, no change
is made to the stabilizers as we already have an eigenstate of the
operator.  For example, if we have two qubits and stabilizers $Z_1$
and $-Z_2$, measuring the $Z_1Z_2$ operator will always give the
$-1$ eigenstate.

If the operator to be measured cannot be expressed as a product of
stabilizers and commutes with each stabilizer, the operator is added
to the list of stabilizers with a sign that depends on whether we
have projected into the $+1$ or $-1$ eigenstate of the operator. For
example, if we have three qubits and stabilizers $Z_1Z_2$ and
$Z_2Z_3$, measuring $Z_2$ will yield one of the $\pm 1$ eigenstates,
meaning we will introduce the new stabilizer  $\pm Z_2$.  Note again
that the probability of the two outcomes is neither recorded nor
known.

Finally, if the operator cannot be expressed as a product of
stabilizers and anticommutes with one or more stabilizers, the
second and subsequent anticommuting stabilizers are multiplied by
the first anticommuting stabilizer to form commuting stabilizers,
and the first anticommuting stabilizer is replaced with the operator
being measured, again with sign depending on which state we have
projected into.  For example, if we have four qubits and stabilizers
$Z_1Z_2$, $Z_2Z_3$ and $Z_3Z_4$, to measure $X_3$ we first multiply
$Z_3Z_4$ by $Z_2Z_3$ and then replace $Z_2Z_3$ with $\pm X_3$ to
give the new set of stabilizers $Z_1Z_2$, $\pm X_3$ and $Z_2Z_4$. In
this instance we know that the probability of the two outcomes is
equal as, given any state $|\psi\rangle$ stabilized by some operator
$S$ and any operator to measure $M$ such that $MS=-SM$, we have
\begin{equation}
|\psi\rangle = \frac{1}{\sqrt{2}}\left[\frac{1}{\sqrt{2}}(1 +
M)|\psi\rangle + \frac{1}{\sqrt{2}}(1 - M)|\psi\rangle\right],
\end{equation}
meaning we have an equal superposition of the $\pm 1$ eigenstates of
$M$.

Many other examples of measurements falling into each of these three
categories will be discussed in subsequent sections.

\section{The surface code}
\label{The surface code}

\begin{figure}
\begin{center}
\resizebox{35mm}{!}{\includegraphics{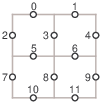}}
\end{center}
\caption{Basic layout of surface code data qubits, each represented
by a circle. A data qubit is located at the center of each edge of a
square lattice. The square lattice is a guide for the eye only, it
does not represent interactions.} \label{surface_code_2x2}
\end{figure}

The surface code was first presented in \cite{Brav98}. A small
surface showing the basic layout of qubits, a square grid with
qubits on each edge, is shown in Fig.~\ref{surface_code_2x2}.  The
stabilizers of this surface are
\begin{eqnarray}
&& \hspace{-5mm} X_0X_2, X_0X_1X_3, X_1X_4, X_2X_5X_7, X_3X_5X_6X_8, \nonumber \\
&& \hspace{-5mm} X_4X_6X_9, X_7X_{10}, X_8X_{10}X_{11}, X_9X_{11} \nonumber \\
&& \hspace{-5mm} Z_0Z_2Z_3Z_5, Z_1Z_3Z_4Z_6, Z_5Z_7Z_8Z_{10},
Z_6Z_8Z_9Z_{11}
\end{eqnarray}
These correspond to a tensor product of $Z$ around each face and $X$
around each vertex.  Note that $X_9X_{11}$ can be expressed as a
product of the other $X$ stabilizers.  This leaves 12 independent
stabilizers on 12 qubits implying a unique state.  Given a $w$ by
$h$ face surface, in general there will be $2wh+w+h$ qubits and
independent stabilizers.

Not shown in Fig.~\ref{surface_code_2x2} are additional qubits on
each face and vertex that enable one to check the sign of the
associated stabilizer. These additional syndrome qubits make the
lattice a simple nearest neighbor connected square lattice.
Discussion of the syndrome qubits and the quantum circuits used to
extract the signs of the stabilizers will be deferred until
Section~\ref{Threshold error rate}.

\begin{figure}
\begin{center}
\resizebox{65mm}{!}{\includegraphics{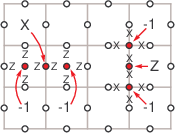}}
\end{center}
\caption{(Color online). Effect of single bit-flips ($X$) and
phase-flips ($Z$) on the surface code --- adjacent face and vertex
stabilizers are made negative, respectively.}
\label{surface_code_single_errors}
\end{figure}

If no errors of any kind occur, the surface remains in the
simultaneous $+1$ eigenstate of every stabilizer.  When discussing
errors, we will restrict our attention to bit-flips and phase-flips.
Very general noise can be tolerated with just the ability to correct
these two types of errors \cite{Shor95}.
Fig.~\ref{surface_code_single_errors} shows the effect of single
bit-flips and phase-flips on the surface code --- the adjacent
stabilizers become negative.  If we could reliably detect when a
stabilizer becomes negative, this clearly would be sufficient to
pinpoint and then correct these single errors.

\begin{figure}
\begin{center}
\resizebox{65mm}{!}{\includegraphics{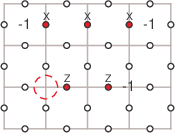}}
\end{center}
\caption{(Color online). Surface code suffering from multiple errors
(indicated by labeled red dots) and an incorrect syndrome
measurement (indicated by the dashed red circle).}
\label{surface_code_multiple_errors}
\end{figure}

Two additional complications need to be accounted for.  Firstly, it
is possible for long chains of errors to occur.  Secondly, it is
possible for the reported eigenvalue of a given stabilizer to be
wrong.  Both of these situations are illustrated in
Fig.~\ref{surface_code_multiple_errors}. To cope with these
complications, we keep track of every time the reported eigenvalue
of each stabilizer changes.  Without loss of generality, let us
focus solely on $Z$ stabilizers, which detect bit-flips, as both
types of errors are treated independently.

\begin{figure}
\begin{center}
\resizebox{85mm}{!}{\includegraphics{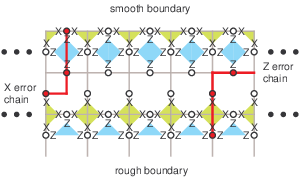}}
\end{center}
\caption{(Color online). Examples of smooth and rough boundaries
including a chain of  $X$ errors ending in a smooth boundary without
changing the sign of any stabilizers, and a chain of $Z$ errors
ending in a rough boundary, also without leaving any evidence of its
presence. $X$/$Z$ stabilizers are represented by green/blue shapes,
respectively.} \label{boundaries}
\end{figure}

\begin{figure}
\begin{center}
\resizebox{85mm}{!}{\includegraphics{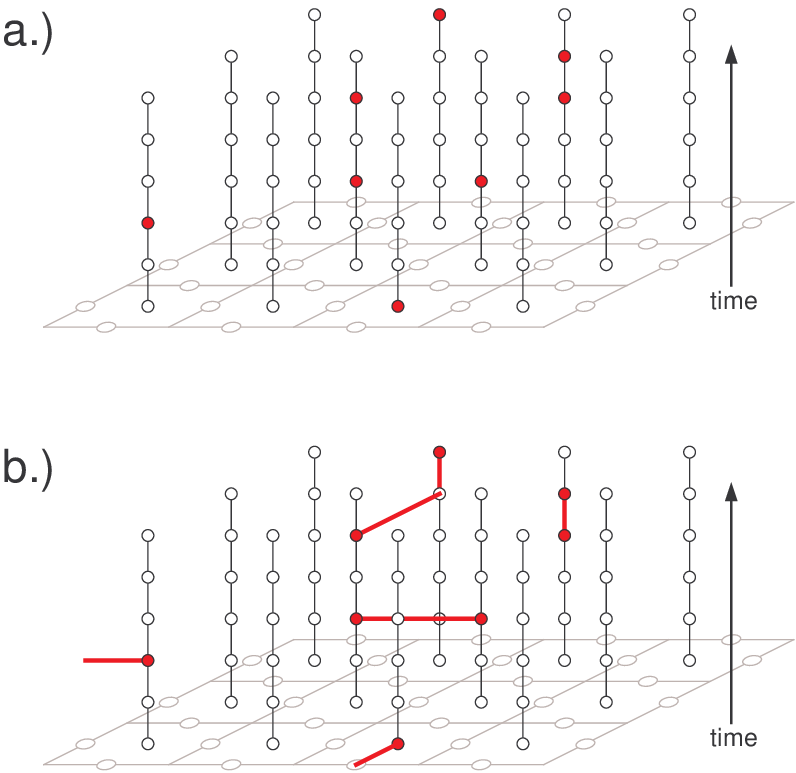}}
\end{center}
\caption{(Color online). a.) Locations in space and time, indicated
by red dots, where and when the reported syndrome is different from
that in the previous time step.  Note that this is not a
three-dimensional physical structure, just a three-dimensional
classical data structure.  b.) Optimal matching highly likely to
lead to a significant reduction of the number of errors if bit-flips
are applied to the spacelike edges.} \label{minimum_weight_2}
\end{figure}

Fig.~\ref{minimum_weight_2}a gives an example of appropriate $Z$
stabilizer information.  In practice, correction is delayed for as
long as possible and pairs of flipped syndromes are then connected
by paths in space and time or ``matched'' such that the total number
of edges used is minimal, as shown in Fig.~\ref{minimum_weight_2}b.
Polynomial time minimum weight matching algorithms exist
\cite{Cook99}, hence this can be done efficiently. Note that $X$
errors can be matched to smooth boundaries and $Z$ errors to rough
boundaries of the surface.  A smooth boundary is a boundary with
four term $Z$ stabilizers and three term $X$ stabilizers as shown in
Fig.~\ref{boundaries}.  A rough boundary is a boundary with four
term $X$ stabilizers and three term $Z$ stabilizers, also shown in
Fig.~\ref{boundaries}. Given a minimum weight matching, bit-flips
are applied to the spacelike edges to correct the errors with high
probability. Further discussion of the details of error correction
will be delayed until Section~\ref{Threshold error rate}.

Initialization of the surface code substrate is not completely
trivial.  If every qubit is prepared in the $|0\rangle$ state, we
automatically have the $+1$ eigenstate of every $Z$ stabilizer, but
when we measure the $X$ stabilizers the eigenstates will be randomly
positive and negative.  For simplicity, we choose to treat the
random negative eigenvalues as errors and correct them.

\section{Logical qubits}
\label{Logical qubits}

\begin{figure}
\begin{center}
\resizebox{50mm}{!}{\includegraphics{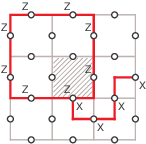}}
\end{center}
\caption{(Color online). Surface code with one additional degree of
freedom introduced by not enforcing the stabilizer associated with
one face (shaded). This face, or a region of such faces, is called a
smooth defect. The degree of freedom can be phase-flipped by any
ring of $Z$ operators encircling the defect and bit-flipped by any
chain of $X$ operators connecting the defect to a smooth boundary.}
\label{surface_code_single_defect}
\end{figure}

Armed with the surface code described in the previous section, we
can now discuss logical qubits.  The simplest logical qubit consists
of a single face where we stop measuring the associated $Z$
stabilizer. This introduces one new degree of freedom into the
surface.  We can manipulate this degree of freedom using any chain
of $X$ operators connecting this face or ``smooth defect'' to a
smooth boundary and any chain of $Z$ operators encircling the smooth
defect as shown in Fig.~\ref{surface_code_single_defect}.  We choose
to call any such $X$ chain $X_L$, and any $Z$ ring $Z_L$. This
implies that, by definition, our logical qubit is initialized to
$|0_L\rangle$ as the surface is initially in the simultaneous $+1$
eigenstate of every $Z$ stabilizer and therefore also in the $+1$
eigenstate of $Z_L$.

\begin{figure}
\begin{center}
\resizebox{65mm}{!}{\includegraphics{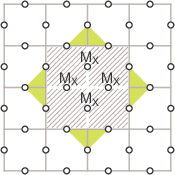}}
\end{center}
\caption{(Color online). Surface code with one degree of freedom
introduced via the measurement $M_X$ of four qubits in the $X$ basis
and removal of five stabilizers.  Note that four new three term $X$
stabilizers are created with not necessarily positive sign
(indicated by green triangles).} \label{surface_code_larger_defect}
\end{figure}

Larger smooth defects can be created using $X$ measurements as shown
in Fig.~\ref{surface_code_larger_defect}.  Note that arbitrarily
large defects still only introduce one degree of freedom.  The given
example shows the removal of four qubits and five stabilizers ---
four $Z$ stabilizers and one $X$ stabilizer. After the $X$
measurements, a number of new three term $X$ stabilizers are created
with not necessarily positive sign.  As in the case of surface
initialization, we will treat any negative eigenvalues as syndrome
changes which will then be matched and corrected with chains of $Z$
operators. The qubits inside the defect, which have been projected
into a product state, play no further role in the computation unless
the defect moves.

\begin{figure}
\begin{center}
\resizebox{85mm}{!}{\includegraphics{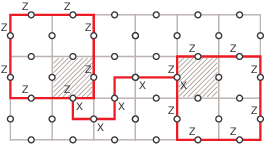}}
\end{center}
\caption{(Color online). Smooth qubit comprised of two smooth
defects. $Z_L$ corresponds to any ring of $Z$ operators around
either defect. $X_L$ corresponds to any chain of $X$ operators
connecting the two defects.} \label{surface_code_2_defects}
\end{figure}

In practice, it is inconvenient to use a logical qubit with a
logical operator that connects to a potentially distant boundary.
This situation can be avoided by using a pair of defects to
represent a single logical qubit as shown in
Fig.~\ref{surface_code_2_defects}. A chain of $X$ operators
connecting the two defects is then used as the $X_L$ operator. The
$Z_L$ operator is any ring of $Z$ operators around either defect
--- these two classes of $Z_L$ operators are equivalent as they have the same
commutation relations.

Effectively, the above means we are choosing to represent an
arbitrary logical state by $\alpha|0_L\rangle|0_L\rangle +
\beta|1_L\rangle|1_L\rangle$ as defined in the opening paragraph of
this section. For the remainder of the review we shall redefine
$|0_L\rangle$ and $|1_L\rangle$ such that an arbitrary logical state
of a double defect logical qubit can be expressed as simply
$\alpha|0_L\rangle + \beta|1_L\rangle$. Note that double smooth
defect logical qubits are also initialized to $|0_L\rangle$ by
default.

\begin{figure}
\begin{center}
\resizebox{85mm}{!}{\includegraphics{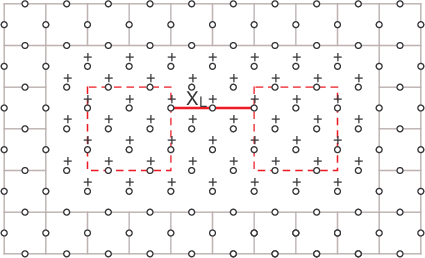}}
\end{center}
\caption{(Color online). Initializing a smooth qubit in the
$|+_L\rangle$ state. After preparing a region of qubits each in the
$|+\rangle$ state, every $X$ stabilizer on the boundary of the
region and every $Z$ stabilizer outside the dashed regions is
measured.  Negative eigenvalues are treated as errors and
corrected.} \label{surface_code_create_smooth_X}
\end{figure}

Double smooth defect logical qubits can also be initialized in the
$|+_L\rangle$ state by first preparing a region of $|+\rangle$ as
shown in Fig.~\ref{surface_code_create_smooth_X}. Such a region is
automatically in the $+1$ eigenstate of $X_L$ operators and $X$
stabilizers not intersecting the boundary.   $X$ stabilizers on the
boundary will have random sign. Smooth defects are then created by
measuring all $Z$ stabilizers outside the desired defect locations.
The signs of the $Z$ stabilizers will be random and we will again
treat negative stabilizers as syndrome changes, match them and
correct them with chains of $X$ operators. We will henceforth refer
to a double smooth defect logical qubit as simply a smooth qubit.

\begin{figure}
\begin{center}
\resizebox{80mm}{!}{\includegraphics{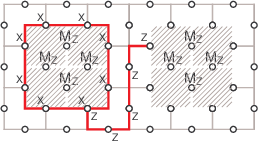}}
\end{center}
\caption{(Color online). Initializing a rough qubit in the
$|+_L\rangle$ state via $Z$ basis measurements $M_Z$ and ignoring
stabilizers (shaded). $X_L$ is any ring of $X$ operators around
either defect. $Z_L$ is any chain of $Z$ operators linking the two
defects.} \label{surface_code_larger_defect_rough}
\end{figure}

Rough qubits are also possible to create via $Z$ measurements as
shown in Fig.~\ref{surface_code_larger_defect_rough}.  In this case
the $Z_L$ operator is any chain of $Z$ operators linking the two
defects, and $X_L$ any ring of $X$ operators around either defect.
Rough qubits are initialized to the $+1$ eigenstate of $X_L$,
$|+_L\rangle$, by default, although $|0_L\rangle$ can be prepared
starting with a region of qubits in the $|0\rangle$ state.

\begin{figure}
\begin{center}
\resizebox{60mm}{!}{\includegraphics{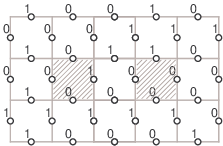}}
\end{center}
\caption{Example of measurement of a smooth qubit in the $Z_L$ basis
in the absence of errors.  Note that the measurements around every
face have even parity whereas the parity of any path of measurements
encircling either defect is odd.  The figure thus corresponds to the
measurement result $|1_L\rangle$.} \label{Z_L_measurement}
\end{figure}

Logical measurement is similar to initialization.  To measure a
smooth qubit in the $Z_L$ basis, a region of qubits encircling
either or both defects is measured in the $Z$ basis.  In the absence
of errors every path encircling either defect will have the same
parity of $Z$ measurements.  If errors are present, they can be
detected and corrected using the standard error correction procedure
as directly measuring qubits in the $Z$ basis is also an acceptable
way to gain information about the eigenvalues of the $Z$ stabilizers
--- even parity of $Z$ measurements around a face
corresponding to a positive eigenvalue and odd parity corresponding
to a negative eigenvalue.  Note that, as shown in
Fig.~\ref{Z_L_measurement}, it is possible for every face to have
even parity, meaning no errors, and every path around either defect
to have odd parity, meaning a readout result of $|1_L\rangle$.

A smooth qubit can be measured in the $X_L$ basis by measuring a
region including both defects in the $X$ basis. In this instance the
parity of all chains of $X$ measurements connecting the two defects
will be the same in the absence of errors.  Similarly, rough qubits
can be easily measured in either logical basis.

\section{Logical CNOT}
\label{Logical CNOT}

So far, we have discussed two types of logical qubits, smooth and
rough, schemes to initialize and measure them in the $Z_L$ and $X_L$
bases, and $Z_L$ and $X_L$ operations.  The only two logical qubit
gate in this scheme is the logical CNOT gate.  To understand how
logical CNOT works, we first need to understand in detail the effect
of moving a smooth defect.

\begin{figure}
\begin{center}
\resizebox{60mm}{!}{\includegraphics{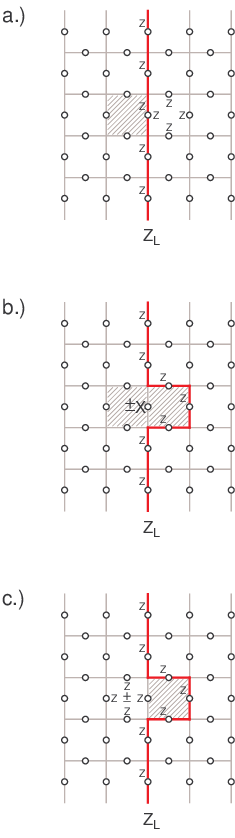}}
\end{center}
\caption{(Color online). a.) Smooth defect and surface in the $+1$
eigenstate of $Z_L$. b.) After measuring the center qubit in the $X$
basis, the shape of the $Z_L$ operator is deformed.  c.) Measuring
and possibly correcting the indicated $Z$ stabilizer using a
bit-flip on the center qubit completes the movement of the defect.}
\label{move_smooth_defect}
\end{figure}

Consider Fig.~\ref{move_smooth_defect}a.  This shows a smooth defect
and two stabilizers
--- a single face $Z$ stabilizer and a $Z_L$ stabilizer.  If
we now measure the center qubit in the $X$ basis as shown in
Fig.~\ref{move_smooth_defect}b, we will be left with the center
qubit in the $\pm X$ eigenstate, and a stabilizer equal to the
product of the face and the path.  We have effectively deformed the
shape of the $Z_L$ stabilizer without changing its sign.  The
movement of the defect can be completed by measuring the $Z$
stabilizer indicated in Fig.~\ref{move_smooth_defect}c, and possibly
correcting the sign of the result by applying an $X$ operator to the
center qubit.  By repeating this process we can see that moving a
smooth defect deforms the shape of $Z_L$ stabilizers passing nearby.

\begin{figure}
\begin{center}
\resizebox{70mm}{!}{\includegraphics{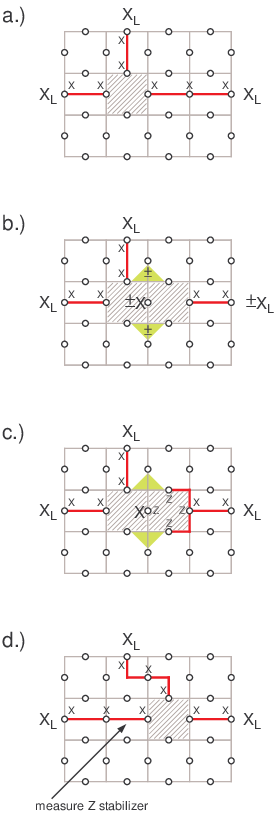}}
\end{center}
\caption{(Color online). a.) Smooth defect and surface in the $+1$
eigenstate of $X_L$. b.) After measuring the center qubit in the $X$
basis, it is possible that three term $X$ stabilizers and $X_L$
stabilizers with negative sign are created (potential locations
indicated in green).  c.) All signs can be corrected by applying the
appropriate single qubit $Z$ operators and chains of $Z$ operators.
d.) Measuring and possibly correcting the indicated $Z$ stabilizer
using a bit-flip on the center qubit completes the movement of the
defect.} \label{move_rough_defect}
\end{figure}

Consider Fig.~\ref{move_rough_defect}a.  This shows a smooth defect
and three $X_L$ stabilizers.  Measuring the center qubit in the $X$
basis as before, we see in Fig.~\ref{move_rough_defect}b that this
has potential side-effects, with a negative eigenvalue indicating
the creation of three term negative $X$ stabilizers and $X_L$
stabilizers of changed sign.  As shown in
Fig.~\ref{move_rough_defect}c, the measured qubit, or qubits in the
case of a larger defect, are individually phase-flipped to ensure
they are all in the $+1$ eigenstate. Pairs of three term negative
$X$ stabilizers are corrected with chains of phase flips along the
boundary of the defect.  Note that this also corrects any $X_L$
stabilizers of changed sign. Fig.~\ref{move_rough_defect}d shows the
effect of completing the movement of the defect by measuring the
appropriate $Z$ stabilizer.  With the signs of the $X_L$ stabilizers
appropriately corrected, all $X_L$ stabilizers attached to the
defect remain attached to the defect with unchanged sign.  By
repeating this process we can see that moving a smooth defect drags
around $X_L$ stabilizers attached to it.

\begin{figure*}
\begin{center}
\resizebox{150mm}{!}{\includegraphics{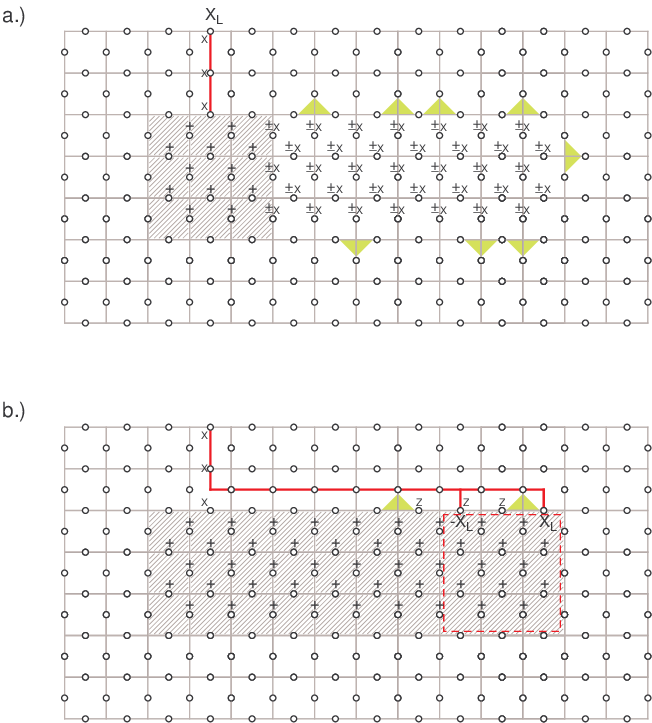}}
\end{center}
\caption{(Color online). a.)  Movement of a large smooth defect via
many measurements in the $X$ basis.  Many pairs of three term $X$
stabilizers with negative sign are likely to be created (indicated
in green). b.) After several rounds of error correction, it becomes
exponentially unlikely that three term $X$ stabilizers with negative
sign remain that were generated in the measurement round. New chains
of errors on the boundary can occur, but these will be corrected
during normal error correction after the size of the defect is
reduced to complete the movement.} \label{move_large_smooth_defect}
\end{figure*}

At first glance, the procedure described in the previous paragraph
does not appear to be fault-tolerant as it seems to rely on perfect
measurement and correction of single qubits.  Indeed, the procedure
is not fault-tolerant unless a larger defect is used as shown in
Fig.~\ref{move_large_smooth_defect}a. After measuring a region of
qubits in the $X$ basis and using the individual measurements to
give the sign of the $X$ stabilizers across the entire measured
region by taking their local parity, we use phase-flips to reset
them to the $+1$ eigenstate as best as we are able.  We do not
assume that we achieve this perfectly. Resetting helps simplify the
later incorporation of these qubits into the surface code.

The three term negative $X$ stabilizers on the boundary left over
after resetting are again treated as syndrome changes and corrected
using the procedure outlined in Section~\ref{The surface code} and
described in more detail in Section~\ref{Threshold error rate} with
the exception that if the procedure suggests connecting to syndrome
changes on the boundary, either new or old, the direction of the
chain of operators correcting these changes is chosen such that the
minimum number of sites on the old (reliable) boundary are changed.
By doing this, every round error correction makes it exponentially
less likely that the three term negative $X$ stabilizers created
during the measurement step in a fixed unit length of boundary still
remain. This implies that the number of rounds of error correction
required to achieve a fixed probability of no errors remaining from
the measurement step only grows logarithmically with the length of
boundary. Note, however, that during the correction procedure new
errors can occur.  The primary desirable feature of these new errors
is that they are unlikely to form very long chains.

Fig.~\ref{move_large_smooth_defect}b shows a potential challenge
when it is time to shrink the size of the defect and complete its
movement.  It is possible for a pair of errors to remain with one
error on the boundary of the region of defect about to be healed,
and the other error on the boundary of the region of defect to
remain.  As shown, without correction, this would result in $X_L$
stabilizers with sign dependent on where they attach to the moved
defect --- a situation that is not allowed.  However, by measuring
all appropriate $X$ stabilizers outside and on the boundary of the
final position of the defect (indicated by a dashed red line) before
measuring all of the $Z$ stabilizers outside the final position, the
presence of these two errors is preserved and subsequent correction
by a chain of $Z$ operators ensures that the sign of all deformed
$X_L$ stabilizers is the same.

To summarize the smooth defect movement procedure, a region of
qubits is measured in the $X$ basis, corrected as best as possible
so that each measured qubit is in the $+1$ eigenstate, which takes
constant time, several rounds of error correction are then applied
until it is sufficiently likely that only new errors occurring after
the initial measurements are now present on the boundary of the new
region, which takes a time that grows logarithmically in the length
of the boundary, then measurement of all $X$ stabilizers outside and
on the boundary of the desired final position of the defect,
measurement of all $Z$ stabilizers outside the final defect
position, and finally error correction until it is sufficiently
likely that only errors occurring after the $Z$ stabilizers were
measured remain, which takes a time that grows logarithmically in
the area being corrected.  This movement procedure deforms nearby
$Z_L$ stabilizers and drags around $X_L$ stabilizers attached to the
defect and takes a total time that grows only logarithmically in the
distance the defect is moved.

\begin{figure*}
\begin{center}
\resizebox{125mm}{!}{\includegraphics{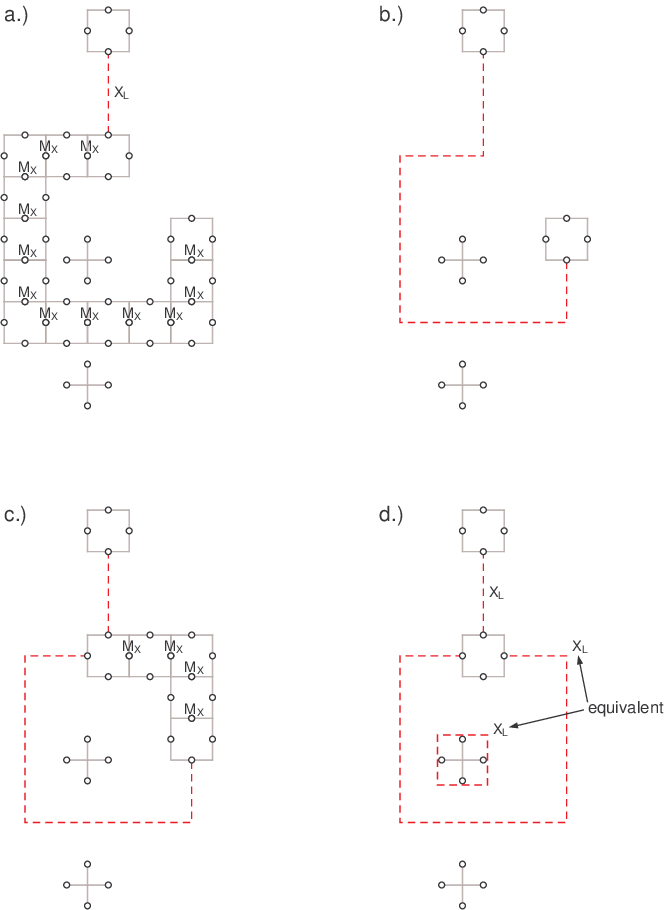}}
\end{center}
\caption{(Color online). a.) Surface containing a smooth qubit in
the $+1$ eigenstate of $X_L$ and a rough qubit.  The lower smooth
defect has been braided around the upper rough defect using $X$
measurements. Note that is not possible to complete the braiding in
one step as a ring of $X$ measurements corresponds to measurement of
the rough qubit in the $X_L$ basis.  b.) Via correction of many $Z$
stabilizers, the $X_L$ operator is dragged around the upper rough
defect. c.) Additional $X$ measurements extend the defect back to
its original position.  d.) Further correction of $Z$ stabilizers
returns the defects to their original positions but the surface is
now in the $+1$ eigenstate of both the smooth and rough $X_L$
operator.} \label{CNOT_stab2_X}
\end{figure*}

\begin{figure*}
\begin{center}
\resizebox{125mm}{!}{\includegraphics{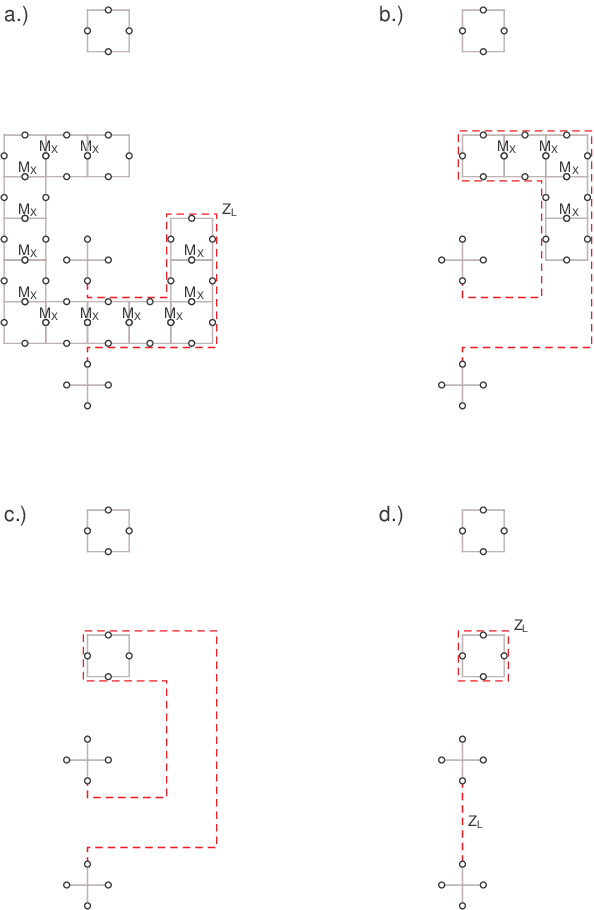}}
\end{center}
\caption{(Color online). a.) Surface containing a smooth defect and
a rough defect in the  $+1$ eigenstate of $Z_L$.  The lower smooth
defect has been braided around the upper rough defect using $X$
measurements, deforming the shape of the rough $Z_L$ operator. b.)
By first correcting many $Z$ stabilizers and then performing further
$X$ measurements, the smooth defect can be extended back to its
original position.  c.) A final round of $Z$ stabilizer correction
returns the defects to their original configuration but with the
state of the surface changed.  d.) The $Z_L$ operator shown in part
c is equivalent to the tensor product of smooth and rough $Z_L$.}
\label{CNOT_stab2_Z}
\end{figure*}

Now that we have a thorough understanding of the effect of moving a
smooth defect, we can return to the question of how to build a
logical CNOT.  Any gate can be completely specified by stating its
action on computational basis states, and can be specified up to
global phase by stating its action on a basis of stabilizers.
Specifically, if we have a system of two qubits and denote the CNOT
between them with the first qubit as the control as $\Lambda_{12}$,
by simple matrix multiplication we can show that the following
relationships hold
\begin{eqnarray}
\Lambda_{12}(I\otimes X)\Lambda_{12}^{\dag}= I\otimes X \label{CNOT1} \\
\Lambda_{12}(X\otimes I)\Lambda_{12}^{\dag}= X\otimes X \label{CNOT2} \\
\Lambda_{12}(I\otimes Z)\Lambda_{12}^{\dag}= Z\otimes Z \label{CNOT3} \\
\Lambda_{12}(Z\otimes I)\Lambda_{12}^{\dag}= Z\otimes I
\label{CNOT4}
\end{eqnarray}
These relationships can be combined to determine the action of CNOT
on an arbitrary two-qubit stabilizer.  To show that we have a
logical CNOT, it is sufficient to show that we can transform logical
stabilizers in the above manner.
Figs.~\ref{CNOT_stab2_X}--\ref{CNOT_stab2_Z} show that the logical
versions of Eq.~{\ref{CNOT2}} and Eq.~{\ref{CNOT3}} hold if we use a
smooth qubit as the control and a rough qubit as the target and
braid one of the smooth defects around one of the rough defects. It
is not important in which direction the braiding is done, only that
the defect return to its initial position.  It is also not important
which smooth defect is moved nor which rough defect it is braided
around.  It is similarly straightforward to show that
Eq.~{\ref{CNOT1}} and Eq.~{\ref{CNOT4}} hold.

\begin{figure}
\begin{center}
\resizebox{85mm}{!}{\includegraphics{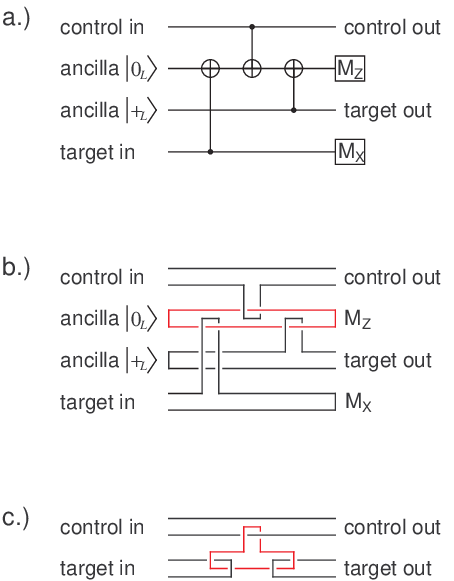}}
\end{center}
\caption{(Color online). Smooth qubits are represented by black
lines, rough qubits by red lines. a.) Circuit equivalent to
$Z^{M_X}$ on the target qubit followed by CNOT between the control
and target qubit followed by $X^{M_Z}$ on the target qubit. b.)
Schematic representing the initialization, braiding and measurement
of defects in a surface code to implement Fig.~\ref{CNOT_smooth}a.
Time runs from left to right, and the surface code should be
imagined oriented vertically and into and out of the page.  c.)
Simplified schematic equivalent to Fig.~\ref{CNOT_smooth}b.}
\label{CNOT_smooth}
\end{figure}

We do not yet have what we truly need --- a CNOT between logical
qubits of the same type.  Define $M_X$, $M_Z$ to take the value 0
when the +1 eigenstate is measured and 1 when the -1 eigenstate is
measured. Consider Fig.~\ref{CNOT_smooth}a.  This is built entirely
out of logical circuit elements described above and is equivalent to
$Z^{M_X}$ on the target qubit followed by CNOT followed by $X^{M_Z}$
on the target qubit.  This is in turn equivalent to CNOT followed by
$(Z\otimes Z)^{M_X}$ followed by $X^{M_Z}$ on the target qubit.  We
will adopt the policy of applying corrective logical operations
based on the measurement results immediately after such a CNOT to
simplify the discussion of more complicated circuits.
Fig.~\ref{CNOT_smooth}a can also be represented as a braiding of
defects of different types in two dimensions of space and one
dimension of time as shown in Fig.~\ref{CNOT_smooth}b and simplified
in Fig.~\ref{CNOT_smooth}c. Note that since logical CNOT is built
out of defect movement, which takes a time that grows
logarithmically in the distance moved, and defect measurement, which
takes constant time, the total time required to execute logical CNOT
grows only logarithmically in the separation of the logical qubits.

\section{State injection and non-Clifford gates}
\label{State injection and non-Clifford gates}

The set of gates discussed so far is not universal.  To complete the
universal set, we will firstly describe how it is possible to
non-fault-tolerantly prepare arbitrary logical states, and then
discuss state distillation \cite{Brav05,Reic05} and non-Clifford
gates based on these distilled states.

\subsection{State injection}
\label{State injection}

\begin{figure}
\begin{center}
\resizebox{50mm}{!}{\includegraphics{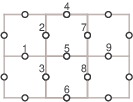}}
\end{center}
\caption{Surface code fragment and numbered qubits used to assist
the visualization of the discussion of Section~\ref{State
injection}.} \label{surface_code_2x3}
\end{figure}

Consider Fig.~\ref{surface_code_2x3}.  We will focus on the numbered
qubits and the four stabilizers $X_1X_2X_3X_5$, $X_5X_7X_8X_9$,
$Z_2Z_4Z_5Z_7$, $Z_3Z_5Z_6Z_8$ centered on qubit 5. The discussion
of this section applies to a surface of arbitrary size
--- we shall see that none of the necessary manipulations affect
stabilizers further away. We shall only explicitly work through the
creation of an arbitrary rough qubit --- the procedure for creating
an arbitrary smooth qubit can be obtained by exchanging the roles of
$X$ and $Z$.

To create an arbitrary rough qubit, begin by measuring qubit 5 in
the $X$ basis.  This gives a state stabilized by
\begin{equation}
\begin{array}{c|ccccccccc}
           & 1 & 2 & 3 & 4 & 5 & 6 & 7 & 8 & 9 \\ \cline{2-10}
           & X & X & X &   & X &   &   &   &   \\
           &   &   &   &   & X &   & X & X & X \\
(-1)^{M_X} &   &   &   &   & X &   &   &   &   \\
           &   & Z & Z & Z &   & Z & Z & Z &   \\
\end{array}
\end{equation}
If the $-1$ eigenstate is obtained, apply either $Z_2Z_4Z_5Z_7$ or
$Z_3Z_5Z_6Z_8$ to create the $+1$ eigenstate
\begin{equation}
\begin{array}{c|ccccccccc}
\textcolor{white}{(-1)^{M_X}}
           & 1 & 2 & 3 & 4 & 5 & 6 & 7 & 8 & 9 \\ \cline{2-10}
           & X & X & X &   &   &   &   &   &   \\
           &   &   &   &   &   &   & X & X & X \\
           &   &   &   &   & X &   &   &   &   \\
           &   & Z & Z & Z &   & Z & Z & Z &   \\
\end{array}
\end{equation}
Next, Hadamard transform (for pedagogical clarity) and then
unitarily rotate qubit 5 to the desired state
\begin{eqnarray}
& \alpha \left( \begin{array}{c|ccccccccc}
           & 1 & 2 & 3 & 4 & 5 & 6 & 7 & 8 & 9 \\ \cline{2-10}
           & X & X & X &   &   &   &   &   &   \\
           &   &   &   &   &   &   & X & X & X \\
           &   &   &   &   & Z &   &   &   &   \\
           &   & Z & Z & Z &   & Z & Z & Z &   \\
\end{array} \right) & \nonumber \\
& + \beta \left( \begin{array}{c|ccccccccc}
           & 1 & 2 & 3 & 4 & 5 & 6 & 7 & 8 & 9 \\ \cline{2-10}
           & X & X & X &   &   &   &   &   &   \\
           &   &   &   &   &   &   & X & X & X \\
           &   &   &   &   &-Z &   &   &   &   \\
           &   & Z & Z & Z &   & Z & Z & Z &   \\
\end{array} \right) &
\end{eqnarray}
Measure either $Z_2Z_4Z_5Z_7$ or $Z_3Z_5Z_6Z_8$
\begin{eqnarray}
& \alpha \left( \begin{array}{c|ccccccccc}
           & 1 & 2 & 3 & 4 & 5 & 6 & 7 & 8 & 9 \\ \cline{2-10}
           & X & X & X &   &   &   & X & X & X \\
           &   &   &   &   & Z &   &   &   &   \\
(-1)^{M_Z} &   & Z &   & Z & Z &   & Z &   &   \\
(-1)^{M_Z} &   &   & Z &   & Z & Z &   & Z &   \\
\end{array} \right) & \nonumber \\
& + \beta \left( \begin{array}{c|ccccccccc}
           & 1 & 2 & 3 & 4 & 5 & 6 & 7 & 8 & 9 \\ \cline{2-10}
           & X & X & X &   &   &   & X & X & X \\
           &   &   &   &   &-Z &   &   &   &   \\
(-1)^{M_Z} &   & Z &   & Z & Z &   & Z &   &   \\
(-1)^{M_Z} &   &   & Z &   & Z & Z &   & Z &   \\
\end{array} \right) &
\end{eqnarray}
If the $-1$ eigenstate of $Z_2Z_4Z_5Z_7$ and $Z_3Z_5Z_6Z_8$ is
obtained, apply $X_5$ and then either $X_1X_2X_3X_5$ or
$X_5X_7X_8X_9$ to give the desired logical state
\begin{eqnarray}
& \alpha \left( \begin{array}{c|ccccccccc}
           & 1 & 2 & 3 & 4 & 5 & 6 & 7 & 8 & 9 \\ \cline{2-10}
           & X & X & X &   &   &   & X & X & X \\
           &   &   &   &   & Z &   &   &   &   \\
           &   & Z &   & Z & Z &   & Z &   &   \\
           &   &   & Z &   & Z & Z &   & Z &   \\
\end{array} \right) & \nonumber \\
& + \beta \left( \begin{array}{c|ccccccccc}
           & 1 & 2 & 3 & 4 & 5 & 6 & 7 & 8 & 9 \\ \cline{2-10}
           & X & X & X &   &   &   & X & X & X \\
           &   &   &   &   &-Z &   &   &   &   \\
           &   & Z &   & Z & Z &   & Z &   &   \\
           &   &   & Z &   & Z & Z &   & Z &   \\
\end{array} \right) &
\end{eqnarray}

After creating an arbitrary logical qubit using the procedure above,
the two halves of the logical qubit would be both moved apart and
made larger as quickly as possible to make the logical qubit
fault-tolerant.

\subsection{State distillation}
\label{State distillation}

\begin{figure}
\begin{center}
\includegraphics[width=0.9\columnwidth]{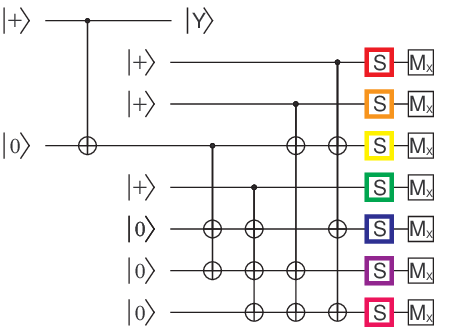}
\end{center}
\caption{Distillation circuit for the $|Y\rangle$ ancilla state. A Bell pair is created, then one qubit from the pair is encoded with 6 ancilla qubits using the Steane code. The seven encoded qubits are then each rotated with an $S$ gate, each of which consumes a $|Y\rangle$ state. The output from each $S$ gate is then measured in the $X$ basis, and the results indicate whether the remaining qubit should be kept. Denoting the measurements bottom to top as $M_0,\ldots,M_6$, with values 0 or 1, the remaining qubit is kept provided the parities of $M_0M_1M_2M_3$, $M_0M_1M_4M_5$, and $M_0M_2M_4M_6$ are even. If the probability of each consumed $|Y\rangle$ state containing a $Z$ error is $p$, the probability of not obtaining these three even parities is $7p$. If the parity of $M_4M_5M_6$ is even, the output is actually $Z|Y\rangle$. The accepted output will have probability of unknown error $7p^3$.}\label{fig.ydistill}
\end{figure}

\begin{figure}
\begin{center}
\includegraphics[width=0.9\columnwidth]{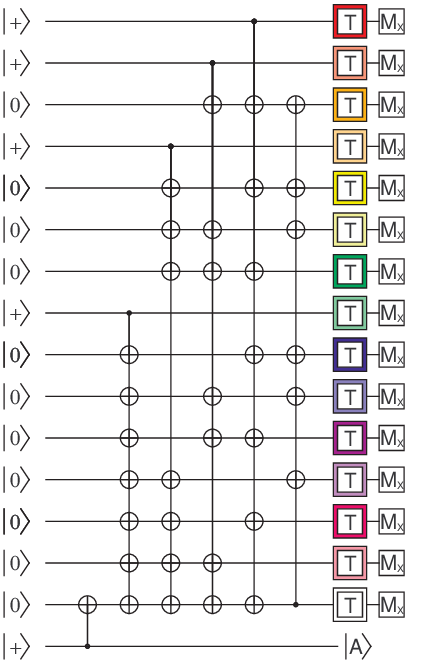}
\end{center}
\caption{Distillation circuit for the $|A\rangle$ ancilla state. A Bell pair is created, and one qubit from the pair encoded with 14 ancilla logical qubits using the Reed-Muller code. The fifteen encoded qubits are then each rotated with a $T$ gate. The ancillae for the $T$ gates are the $|A\rangle$ states that are being purified, prepared either by state injection or produced in a previous distillation round. Following the $T$ gates, fifteen $X$ basis measurements $M_X$ are made. Denoting the measurements bottom to top as $M_0,\ldots,M_{14}$, with values 0 or 1, the remaining qubit is kept provided the parities of $M_0M_1M_2M_3M_4M_5M_6M_7$, $M_0M_1M_2M_3M_8M_9M_{10}M_{11}$, $M_0M_1M_4M_5M_8M_9M_{12}M_{13}$, and $M_0M_2M_4M_6M_8M_{10}M_{12}M_{14}$ are even. If the probability of each consumed $|A\rangle$ state containing a $Z$ error is $p$, the probability of not obtaining these four even parities is $15p$. If the parity of $M_8M_9M_{10}M_{11}M_{12}M_{13}M_{14}$ is even (odd), the output is $X|A\rangle$ ($ZX|A\rangle$). The accepted output will have probability of unknown error $35p^3$.}\label{fig.adistill}
\end{figure}

For our purposes, we are interested in the injection of two particular states $|Y\rangle = |0\rangle + i|1\rangle$ and $|A\rangle = |0\rangle + e^{i\pi/4}|1\rangle$. These states can be made arbitrarily precise through repeated state distillation. The state distillation processes we consider here are probabilistic, taking multiple imperfect input states and producing a single better output state. Circuits for distilling the $|Y\rangle$ and $|A\rangle$ ancilla states are shown in Figs.~\ref{fig.ydistill} and \ref{fig.adistill}. The figure captions contain descriptions of the structure and operation of these circuits.

Note that both Fig.~\ref{fig.ydistill} and \ref{fig.adistill} are
made of operations that can be performed easily and efficiently
using the surface code.  In particular, the single control multiple
target CNOTs can be implemented in the same amount of time as a
single CNOT.  The input $|Y\rangle$ and $|A\rangle$ states would be
created factory style, with any errors detected early in the
non-fault-tolerant process of their creation resulting in a restart
of the creation process.  Logical ancilla states that are likely to
be sufficiently good would then be recursively fed into logical
surface code versions of Fig.~\ref{fig.ydistill} and \ref{fig.adistill} until sufficiently high fidelity ancilla
states are obtained.

See \cite{Fowl12h,Brav12b,Jone12} for more advanced state distillation methods.

\subsection{Non-Clifford gates}
\label{Non-Clifford gates}

\begin{figure}
\begin{center}
\resizebox{75mm}{!}{\includegraphics{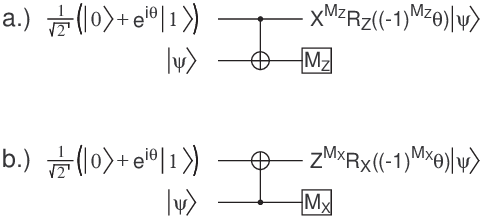}}
\end{center}
\caption{a.) Circuit performing the single qubit unitary
$X^{M_Z}R_Z((-1)^{M_Z}\theta)$ given an appropriate ancilla state.
b.) Circuit performing the single qubit unitary
$Z^{M_X}R_X((-1)^{M_X}\theta)$ given an appropriate ancilla state.}
\label{non-Clifford}
\end{figure}

Given states of the form $(|0\rangle +
e^{i\theta}|1\rangle)/\sqrt{2}$, rotations $R_{Z}(\theta)$ and
$R_{X}(\theta)$ can be performed using the circuits shown in
Fig.~\ref{non-Clifford}a and Fig.~\ref{non-Clifford}b respectively.
Note that both of these circuits are probabilistic, and actually
perform rotations $XR_{Z}(-\theta)$ and $ZR_{X}(-\theta)$ if the
measurement indicates a negative eigenstate.  If we wish to apply
$R_Z(\pi/2)$ or $R_X(\pi/2)$ and discover we have actually applied
rotations $XR_Z(-\pi/2)$ or $ZR_X(-\pi/2)$, the correct gate can be
achieved simply by a subsequent application of $Z$ and $X$. If
attempting $R_Z(\pi/4)$ and we discover we have applied
$XR_Z(-\pi/4)$, an ancilla state $|0\rangle + i|1\rangle$ needs to
be ready for an attempt to apply $R_Z(\pi/2)X$. If we again measure
a negative eigenstate, subsequent application of $ZX$ gives the
desired rotation.

\section{Logical Hadamard}
\label{Logical Hadamard}

The Hadamard gate is called for in many quantum algorithms.  In
principle we could simply use the relation
\begin{equation}
H \equiv R_Z(\pi/4)R_X(\pi/4)R_Z(\pi/4) \label{Heq}
\end{equation}
and the constructions of Section~\ref{State injection and
non-Clifford gates}. There is, however, a much more efficient way
\cite{Bomb07}.

\begin{figure}
\begin{center}
\resizebox{85mm}{!}{\includegraphics{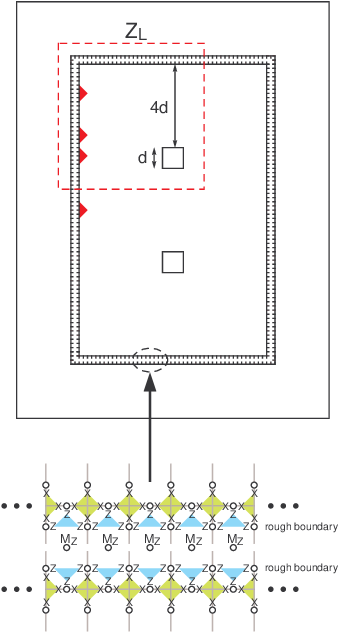}}
\end{center}
\caption{(Color online). A smooth qubit isolated from a larger piece
of surface code using a ring of $Z$ measurements so that logical
Hadamard can be applied by local, transversal Hadamard gates.  Red
triangles represent three term $Z$ stabilizers of negative sign. See
text for details.} \label{Hadamard}
\end{figure}

Consider Fig.~\ref{Hadamard}. This shows a smooth qubit cut out of a
larger lattice using $Z$ measurements.  Note that the three term $Z$
stabilizers thus created would need to be corrected as they would
have random sign.  Without correction, the indicated $Z_L$
stabilizer would have random sign after the measurements. Note that
the ring of $Z$ measurements provides no information about the state
of the smooth qubit --- such a ring is equivalent to the logical
identity operator.

The logical Hadamard gate can now be performed transversely.  Every
face $Z$ stabilizer becomes a vertex $X$ stabilizer.  The rough
boundary becomes a smooth boundary.  The smooth qubit becomes a
rough qubit.  Stabilizers $Z_L$ and $X_L$ are interchanged. This
last point is precisely the action of logical Hadamard.

The interchanging of faces and vertices does create a slight problem
--- faces and vertices are no longer where they should be.  Before
connecting the logical qubit to the rest of the lattice, it would
need to be moved diagonally in any direction a half lattice spacing.
This could be achieved via physical swap gates.  After realignment,
the complete surface of stabilizers would be measured and corrected
once more.

We are still not quite done --- our logical qubit has been converted
from smooth to rough.  By preparing a smooth ancilla qubit in the
$|+_L\rangle$ state and performing a simple smooth-rough CNOT
followed by measurement of the rough qubit in the $Z_L$ basis and
application of $X_L$ if the $-1$ eigenstate is obtained, we can
convert the rough qubit back into a smooth qubit, completing the
process. While not completely trivial, this complete process is
vastly simpler than the necessary ancilla state preparation and
distillation associated with Eq.~(\ref{Heq}).

\section{Threshold error rate}
\label{Threshold error rate}

In our simulations we look at a planar square lattice with two
smooth and two rough boundaries. This type of a lattice lets us
encode one logical qubit.  Our general calculation strategy involves
preparing the system in the simultaneous $+1$ eigenvalue of all $Z$
and $X$ stabilizers and observing how long it takes for the encoded
logical state to change as a result of randomly generating errors.

\begin{figure}
\begin{center}
\resizebox{60mm}{!}{\includegraphics{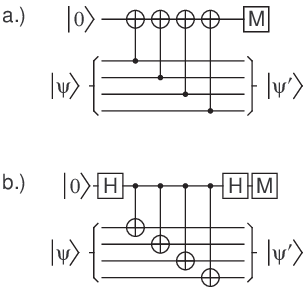}}
\end{center}
\caption{Circuit showing how an additional syndrome qubit (top line
of each figure) is used to measure a.) $Z$ stabilizers, b.) $X$
stabilizers.} \label{syndrome_circuits}
\end{figure}

\begin{figure}
\begin{center}
\resizebox{80mm}{!}{\includegraphics{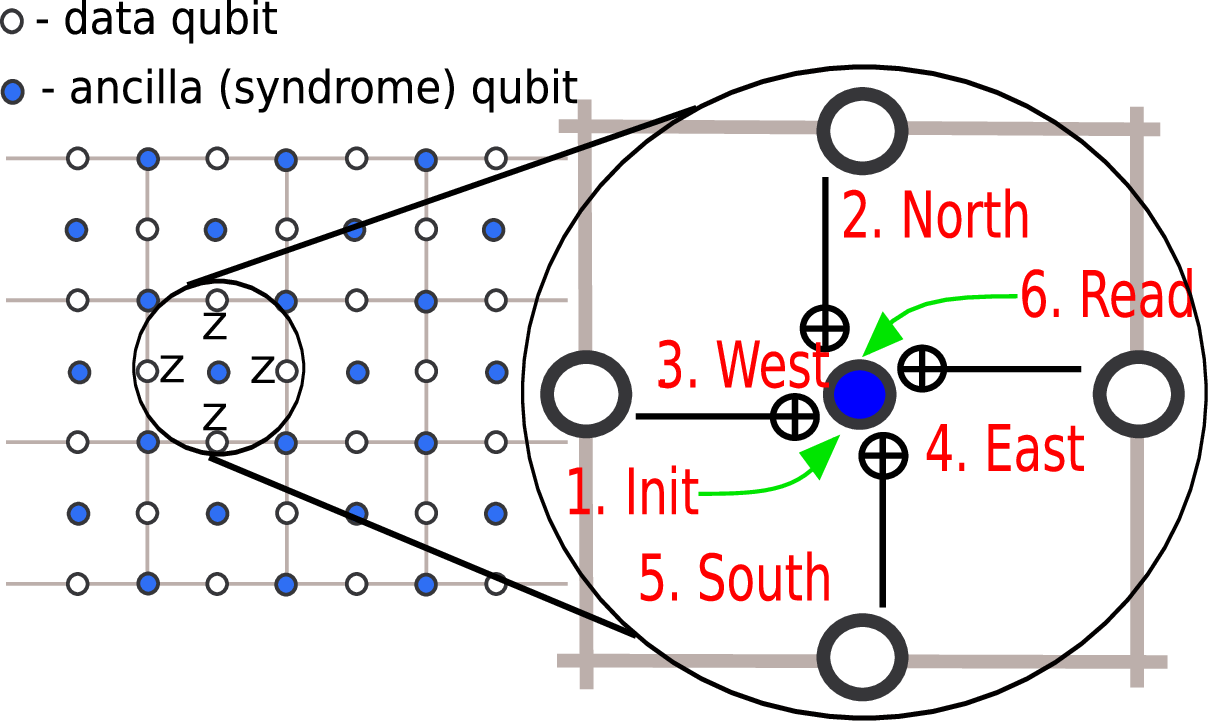}}
\end{center}
\caption{(Color online). Syndrome measurement typically involves six
gates: syndrome initialization, CNOTs with the four surrounding data
qubits (fewer on boundaries) and finally syndrome readout.}
\label{syndrome_readout_diagram}
\end{figure}

We have not yet discussed how $X$ and $Z$ stabilizers are actually
measured.  Fig.~\ref{syndrome_circuits} shows that a fifth syndrome
qubit is required to detect whether the state of the surface
$|\psi\rangle$ is in the $\pm 1$ eigenstate of a $Z$ or $X$
stabilizer \cite{Denn02}. If the surface $|\psi\rangle$ is in
neither eigenstate, the circuit projects the surface into a state
$|\psi'\rangle$ that is one of the $\pm 1$ eigenstates. It takes six
steps to perform such a measurement. The syndromes are initialized,
a CNOT operation is applied between each syndrome and the qubit to
the north, west, east and south, and finally every syndrome qubit is
itself measured, as schematically shown in
Fig.~\ref{syndrome_readout_diagram}. This order of the CNOT gates
has been chosen to ensure that adjacent syndrome circuits sharing a
pair of data qubits are strictly ordered
--- one syndrome circuit touches both data qubits before the other
syndrome circuit.  CNOT gate orders without this property resulted
in entangled syndrome qubits that provide no useful information for
error correction. The placement of a syndrome qubit on each vertex
and in the center of each face, plus the CNOTs required during the
circuit imply that we need a 2-D nearest neighbor coupled lattice of
qubits.

The threshold error rate is derived from four error rates in our
simulations
--- initialization error $p_i$, readout error $p_r$, memory
error $p_m$ and the error associated with a two-qubit gate $p_g$.
Note that we combine any single-qubit gates with neighboring
two-qubit gates and thus do not have a separate single-qubit error
rate.  All four of these error rates are set to the same value $p$
and all operations are assumed to take the same amount of time to
permit our threshold error rate to be compared with others in the
literature \cite{Raus06,Raus07d,Wang09}.

By initialization, we mean initialization to the state $|0\rangle$.
An initialization error is therefore accidental preparation of state
$|1\rangle$ with probability $p$.  By readout, we mean readout in
the $Z$ basis.  A readout error is a classical error --- the qubit
is projected into the $\pm 1$ eigenstate of $Z$, but with
probability $p$ the eigenstate reported by the measurement device is
incorrect.  A memory error is the application of $X$, $Y$ or $Z$,
each with probability $p/3$, to an idle qubit.  A two-qubit gate
error is the application of one of the 15 nontrivial tensor products
of $I$, $X$, $Y$ and $Z$, each with probability $p/15$, after
perfect application of the two-qubit gate.

As was briefly outlined in Section \ref{The surface code}, after
each syndrome is read, its value is checked against a result from
the previous iteration, and if the values differ, the syndrome
change location (in time and space) is recorded. Next, a matching of
all the syndrome changes collected up to this point (an example is
shown in Fig~\ref{minimum_weight_2}a) is used to guess where errors
occurred. Since shorter error chains are more likely than longer
ones, we use a minimum weight matching algorithm to do this
\cite{Cook99}. Before the matching algorithm can find a minimum
weight solution, we convert all the syndrome change results into a
graph, with locations of the syndrome changes representing the
nodes, and edges between these nodes having a weight which depends
on the distance between them. The edge weight is measured in faces
along the special dimensions and syndrome extraction cycles along
the time dimension.

We once again stress that some error chains may begin at the
boundary and end somewhere inside the lattice (see Fig
\ref{boundaries}). In such cases, we can only observe the syndrome
change on the interior of the lattice. To account for this (meaning
enable the matching algorithm to guess that the error chain started
on a boundary), for every interior node, we always create a closest
boundary node. The edges between different boundary nodes are set to
be of weight zero. One way to prepare our graph would be to include
an edge between every pair of nodes (since in principle we don't
know where actual errors occurred), but in practice this is not
necessary. Only edges that connect nodes which are not further from
each other than the sum of the weights between each node and their
closest boundary nodes are included since nodes that are further
apart will always be matched with their respective closest
boundaries in preference to each other.

We further optimize graph creation by noting that matches which are
temporally far behind the current time step are unlikely to be
modified by recent syndrome changes and therefore can be
``remembered'' from previous iterations. These techniques let us
minimize the size of the graph that is passed to the matching
algorithm which, despite scaling polynomially in the number of edges
and nodes, can often still take substantial computing time. An
example of a successful minimum weight perfect match is shown in
Fig~\ref{minimum_weight_2}b.

\begin{figure}
\begin{center}
\resizebox{60mm}{!}{\includegraphics{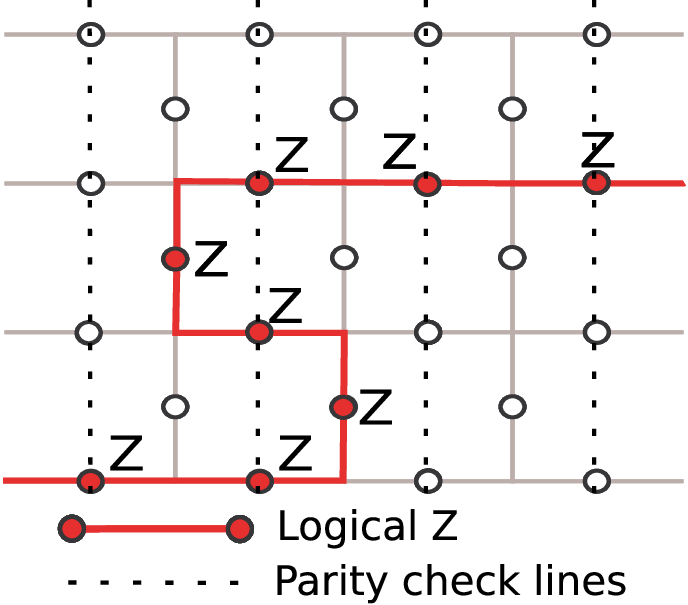}}
\end{center}
\caption{(Color online). Logical $Z$ ($X$) error detection involves
checking if the parity of $Z$ ($X$) operators along any of the
vertical (horizontal) lines of qubits is odd. Above, we show this in
an example of a logical $Z$
error.}\label{fig:logical_error_detection}
\end{figure}

As outlined at the beginning of this section, in order to know if
the simulation should continue or not, we need to determine whether
the lattice suffered a logical error (and hence the encoded state
has changed). A logical error corresponds to a chain of errors that
starts on one boundary and ends on the opposite one. In order to
detect if a logical error has occurred, we repeat the readout cycle
with all the error sources set to zero (i.e. set
$p_{i}=p_{r}=p_{m}=p_{g}=0$, in other words have a ``perfect
readout''). This allows us to be certain that any logical $Z$ ($X$)
error can be recognized by solely checking if the parity of $Z$
($X$) operators crossing a vertical (horizontal) line of qubits is
odd. A simple example of this is shown in Fig
\ref{fig:logical_error_detection}. If no logical error is detected,
we revert the simulation state to what it was just before the
``perfect readout'' cycle was executed and continue on.

\begin{figure}
\begin{center}
\resizebox{80mm}{!}{\includegraphics{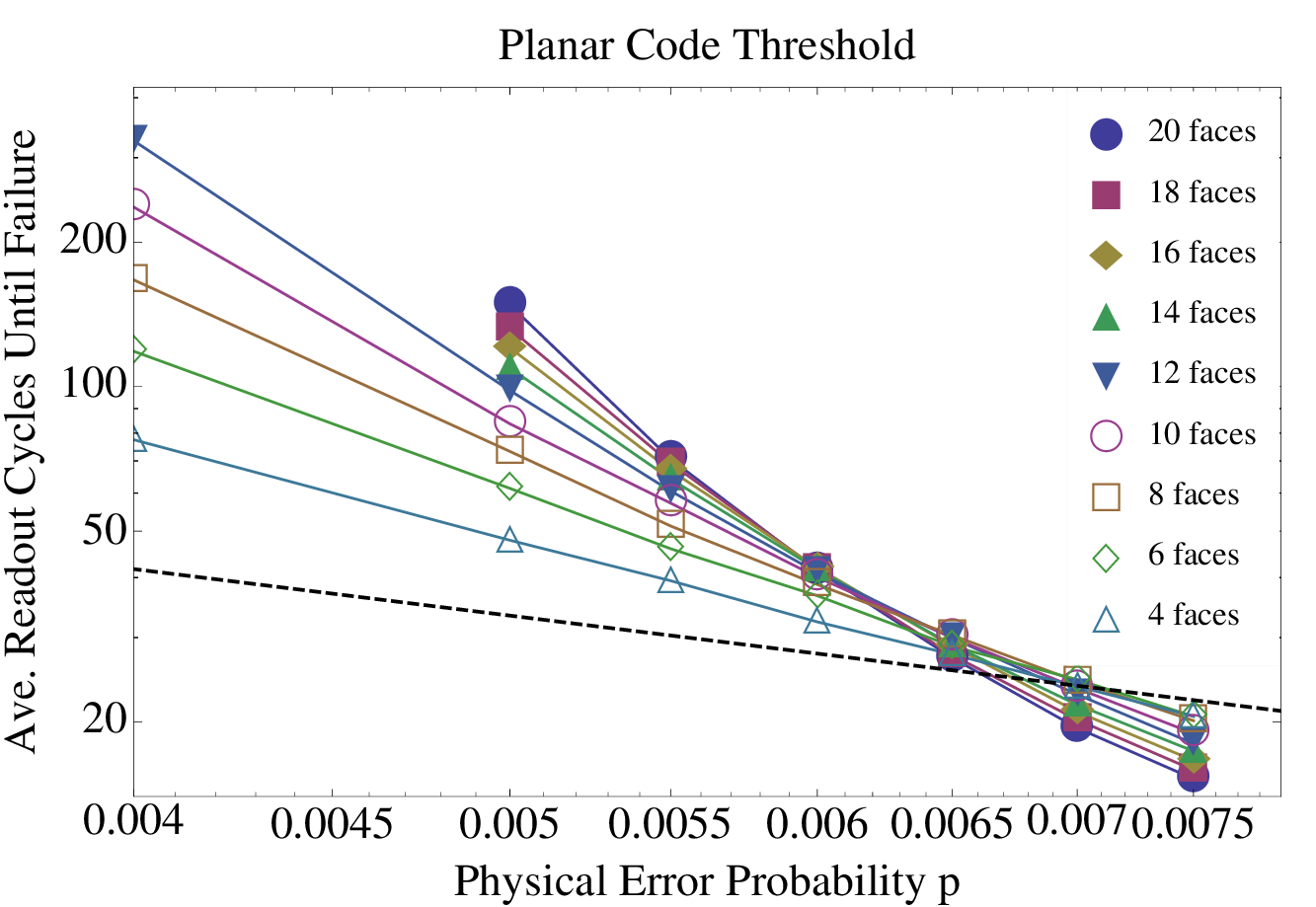}}
\end{center}
\caption{(Color online). A plot of average time until failure versus
the physical error rate $p$. A threshold is observed at $p \approx
6.0 \times 10^{-3}$ where the curves of different lattice sizes
cross. The case where no error correction is used (single qubit) is
represented by a dashed line.}\label{threshold_results}
\end{figure}

During every run, we note how many syndrome extraction cycles it
took for a logical error to be observed. The simulation is repeated
many times for different lattice sizes and values of the physical
error rate $p$. All this data is then used to calculate the average
number of steps until a logical error occurs for a given lattice
size and $p$. The graph of Fig~\ref{threshold_results} shows the
obtained results. In it we see a log-log plot of the average time
until failure versus the physical error rate $p$ for lattice sizes
ranging from $4$ to $20$ faces across. We observe a crossing at
approximately $p \approx 6.0 \times 10^{-3}$, which is our numerical
threshold. If the physical error rate is below this threshold value,
the average number of readout cycles until failure can be increased
arbitrarily by increasing the distance of the code (lattice size).

\section{Distributed computing}
\label{Distributed computing}

In this section, we show that distributed quantum computing can be
performed in a natural manner.  For our purposes, a distributed
quantum computer will consist of a number of separate rectangular
lattices of qubits each capable of holding at least two logical
qubits.  Computing shall proceed by first moving logical qubits that
need to interact onto a common plate before attempting the logical
interaction.  The movement of logical qubits from one plate to
another is the only additional capability we need to discuss.

\begin{figure}
\begin{center}
\resizebox{80mm}{!}{\includegraphics{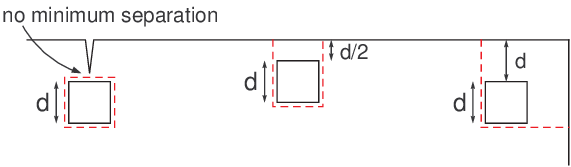}}
\end{center}
\caption{(Color online). Potential $X$ error chains (dashed lines)
around a rough defect and consequent minimum separation from long
straight smooth boundaries and two types of corners such that $X$
error chains beginning and ending on the smooth boundary are not
more likely than an $X$ error ring around the rough defect.}
\label{separation}
\end{figure}

\begin{figure*}
\begin{center}
\resizebox{135mm}{!}{\includegraphics{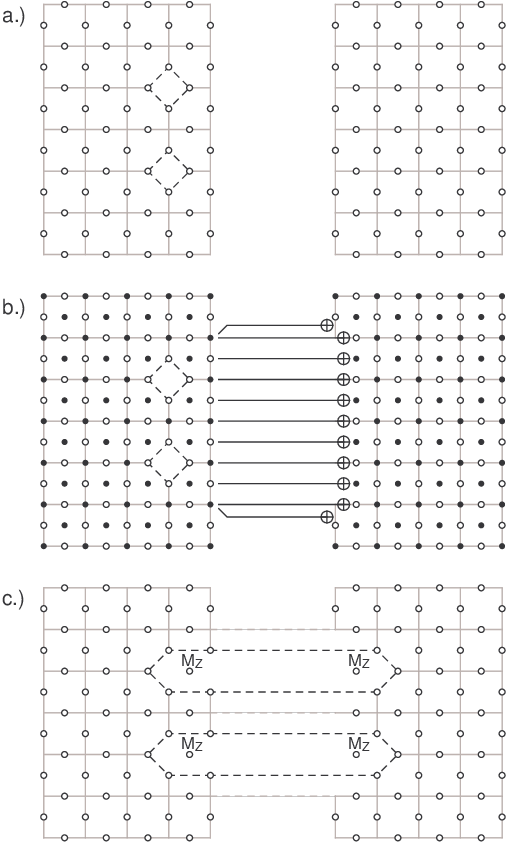}}
\end{center}
\caption{a.) A rough qubit ready to be sent to a separate piece of
surface code.  b.) Remote gates are used to join the two surfaces
together.  c.) A sequence of measurements is used to move the rough
qubit.  After the necessary correction associated with completing
the movement, the long-range gates can be discontinued to separate
the two pieces of surface code once more.} \label{distributed}
\end{figure*}

Consider Fig.~\ref{distributed}a.  This shows a plate containing a
rough qubit and an empty plate.  Note that rough defects do not need
to be kept very well separated from smooth boundaries as no error
chain can link a rough defect with a smooth boundary.
Fig.~\ref{separation} shows the minimum permissible separation from
long straight boundaries and corners. Rough defects do, however,
still need to be kept well separated from each other.

To move the rough qubit from one plate to the other, it must be
possible to perform remote gates between either two complete edges,
or a smaller section of two edges if the plates are large relative
to the size of a logical qubit. Generally speaking, implementing
remote gates would be expected to involve entanglement distribution
and purification \cite{Benn96b}.  We will not discuss the details
here besides mentioning that this leads to significant qubit and
gate overhead implying remote gates should be kept to a minimum.

Consider Fig.~\ref{distributed}b.  This shows the pairs of qubits,
including syndrome qubits, that need to be remotely interacted to
enable a single round of the error correction to proceed seamlessly
across the two plates.  Note that one column of qubits on the empty
plate has been omitted as though it is idle, but note that this
figure does not include the qubits required for entanglement
purification and it is unlikely there would be idle qubits on the
boundary in practice. In general, the joined plates will be in
random eigenstate of both the $X$ and $Z$ stabilizers straddling
both plates.  We shall treat these random values as errors and
correct them.

After correction of the join, the rough qubit can be moved over to
the other plate via $Z$ measurements as shown in
Fig.~\ref{distributed}c.  First the border $Z$ stabilizers of this
extended defect would need to be corrected as discussed in
Section~\ref{Logical CNOT}, then, when shrinking the size of the
defects to move the logical qubit, the unneeded regions of $X$
stabilizers measured and corrected once more.  Both of these
correction procedures take a number of time steps that only grows
logarithmically with the size of the computation and the length or
area being corrected. After the necessary correction has been
completed, error correction can continue on each plate individually
without any further long-range interactions.

The most common reason to move a logical qubit from one plate to
another would be to perform a remote CNOT.  This would be achieved
by creating a rough qubit on the control plate, braiding it around
the control qubit, sending the rough qubit to the target plate and
completing the necessary braiding and measurement operations
entirely on the target plate.

\section{Conclusion and further reading}
\label{Conclusion and further reading}

We have presented a simplified yet comprehensive review of the 2-D
version of the quantum computation scheme originally presented in
\cite{Raus07,Raus07d}. We started with a description of the surface
code, as well as the stabilizer formalism which is used throughout
this paper.  We discussed in detail logical state initialization,
logical CNOT and non-Clifford group gates, which make use of state
distillation. We calculated a numerical threshold for the surface
code and obtained a value of $p \approx 6.0 \times 10^{-3}$ which is
commensurate with other calculations in the literature
\cite{Raus06,Raus07d,Wang09}.

\section{Acknowledgements}

We are much indebted to Robert Raussendorf for extensive and
illuminating discussions.  AGF and AMS acknowledge support from the
Australian Research Council, the Australian Government, and the US
National Security Agency (NSA) and the Army Research Office (ARO)
under contract number W911NF-08-1-0527.

\bibliography{../../../../References} 

\end{document}